\documentstyle[11pt,newpasp,twoside]{article}
\def\edcomment#1{\iffalse\marginpar{\raggedright\sl#1\/}\else\relax\fi}
\reversemarginpar

\begin{document}
\title{General Relativistic Spectra from Accretion Disks around Rapidly Rotating
Neutron Stars}

\author{Sudip Bhattacharyya}

\affil{Indian Institute of Astrophysics, Bangalore 560034, India;
sudip@physics.iisc.ernet.in} 

\author{Dipankar Bhattacharya}

\affil{Raman Research Institute, Bangalore 560012, India}

\author{Ranjeev Misra}

\affil{Department of Physics and Astronomy, Northwestern University, USA}

\author{Arun V. Thampan}

\affil{IUCAA, Pune 411 007, India}

\begin{abstract}

We compute spectra from accretion disks around rapidly rotating neutron stars.
The full effect of general relativity is considered 
for the structure calculation of the stars. We take into
account the Doppler shift, gravitational redshift and light-bending effects 
in order to compute the observed spectra. To facilitate direct comparison
with observations, a simple empirical function is presented which describes
the numerically computed spectra well. This function can in 
principle be used to distinguish between the Newtonian spectra and the 
relativistic spectra. We also discuss the possibility of constraining neutron 
star's equation of state using our spectral models.

\end{abstract}

\section{Introduction}

A large number of low mass X-ray binaries (LMXB) are believed to harbor
neutron stars, rotating rapidly due to accretion-induced angular 
momentum transfer. These systems show many complex spectral and temporal
behaviors. One of the main purpose for studying such behaviors is to understand
the properties of very high density $(\sim 10^{15}$ g cm$^{-3})$ matter at the
neutron star core. Such high densities can not be created in the laboratory 
and only the study of these sources can give a possible answer to this
fundamental question of physics. Here we calculate the equation of state (EOS)
dependent model spectra of the accretion disks around rapidly rotating neutron
stars. These models, when fitted to the observed spectra, can in principle 
help to constrain EOS models 
and hence to understand the properties of core-matter of neutron stars.

\section{Formalism and Results}

We compute the structure of a rapidly rotating neutron star
for realistic EOS models, gravitational masses and rotational speeds 
using the same procedure as Cook, Shapiro, \& Teukolsky (1994).
We use the axisymmetric metric 
(see Bhattacharyya et al. 2000 for description)
\begin{eqnarray}
dS^2 & = & -e^{\rm {\gamma + \rho}} dt^2 + e^{\rm {2\alpha}} 
(d{\bar r}^2 + {\bar r}^2 d {\theta}^2) 
 + e^{\rm {\gamma - \rho}} {\bar r}^2 \sin^2\theta 
 {(d\phi - \omega dt)}^2. 
\end{eqnarray}

\noindent 
To calculate the metric coefficients and the bulk structure parameters
of the neutron star, we solve Einstein's field equations and the equation of
hydrostatic equilibrium simultaneously. For a thin blackbody disk, we compute
the temperature profile (see Bhattacharyya et al. 2000) and hence the 
spectrum (see Bhattacharyya, Bhattacharya, \& Thampan 2001b) 
considering the effect of the 
Doppler shift, gravitational redshift and light-bending effects. We calculate 
the spectra for different EOS models, which therefore gives a way to constrain
EOSs when fitted to the observed spectra. However, the computation of 
the complete spectrum in this manner is rather time-consuming and therefore not
quite suitable for routine use. Therefore, in order to make
our results available for routine spectral 
fitting work, we intend to present a series of approximate
parametric fits to our computed spectra in a forthcoming publication.
As a first step towards this aim, here we present an analytical 
function which describes both the relativistic spectra and the
Newtonian spectra well. This function with three free parameters $(S_o, 
\beta$ and $E_a)$ is given by (see Bhattacharyya, Misra, \& Thampan 2001a for
details)
\begin{eqnarray}
S_{\rm f} ( E ) & = & S_o E_a^{-2/3} ({E\over E_a})^{\gamma} 
exp(-{E\over E_a}),
\end{eqnarray}
\noindent where, $\gamma = -(2/3)(1+E\beta/E_a)$, $E$ is the energy of 
the photons in keV and $S_{\rm f}(E) $ is in units of photons/sec/cm$^2$/keV.
The value of $\beta$-parameter is $\approx 0.4$ for the Newtonian case, 
while it ranges from $0.1$ to $0.35$ for the relativistic cases (if the value 
of inclination angle is not too high). Therefore, constraining this parameter 
by fits to future observational data may indicate
the effect of strong gravity in the observed spectrum.

\section{Conclusion}

This work is a step forward towards constraining EOS models of neutron stars
by fitting the observed spectra of LMXBs.

\begin{quote}
\verb"Bhattacharyya, S., Bhattacharya, D., & Thampan, A. V. 2001b,"\\
\verb"MNRAS, 325, 989"

\verb"Bhattacharyya, S., Misra, R., & Thampan, A. V. 2001a, ApJ,"\\ 
\verb"550, 1"

\verb"Bhattacharyya, S., Thampan, A. V., Misra, R., & Datta, B."\\ 
\verb"2000, ApJ, 542, 473"

\verb"Cook, G. B., Shapiro, S. L., & Teukolsky, S. A. 1994, ApJ,"\\ 
\verb"424, 823"
\end{quote}

\end{document}